\documentclass[manuscript,nonacm]{acmart}

\setcopyright{none}

\usepackage{multirow}
\usepackage{booktabs}
\usepackage{geometry}
\usepackage{array}
\usepackage{caption}
\usepackage{makecell}

\acmConference[CHIWORK '25]{Navigating Generative AI Disclosure, Ownership, and Accountability in Co-Creative Domains Workshop at CHIWORK 2025}{June 23--25,
  2025}{Amsterdam, The Netherlands}

\acmISBN{978-1-4503-XXXX-X/2018/06}

\usepackage{xcolor}

\begin{document}

% Mina: What do you think about the new title? I tried to come up with something shorter and less jargon-y.
% \title{Investigating Factors Influencing Writers' Decisions and Readers' Expectations Around LLM Disclosure in Human-LLM Collaborative Writing}
\title{What Shapes Writers' Decisions to 
Disclose AI Use?}

\author{Jingchao Fang}
\email{jcfang@uchicago.edu}
\orcid{0000-0002-9412-4244}
\affiliation{
  \institution{University of Chicago}
  \city{Chicago}
  \state{Illinois}
  \country{United States}}

\author{Mina Lee}
\email{mnlee@uchicago.edu}
\orcid{0000-0002-0428-4720}
\affiliation{
  \institution{University of Chicago}
  \city{Chicago}
  \state{Illinois}
  \country{United States}}

\renewcommand{\shortauthors}{Fang and Lee}

\begin{abstract}

Have you ever read a blog or social media post and suspected that it was written---at least in part---by artificial intelligence (AI)? 
While transparently acknowledging contributors to writing is generally valued, why some writers choose to disclose or withhold AI involvement remains unclear. 
In this work, we ask \emph{what factors shape writers' decisions to disclose their AI use} as a starting point to effectively advocate for transparency. To shed light on this question, we synthesize
study findings and theoretical frameworks in human-AI interaction and behavioral science. Concretely, we identify and curate a list of factors that could affect writers' decisions regarding disclosure for human-AI co-created content. 
% Moving forward, we plan to investigate these factors comprehensively \mina{unclear what ``comprehensively'' means here} to gain a deeper understanding of their combined effects. \mina{I suggest that we cut the last sentence for completeness for the abstract}
% We expect that taking an integrative approach  
% to provide insights into the potential misalignment between practices and expectations of disclosing LLM use and lay the groundwork for designing interventions to promote greater transparency. 
% \mina{for simplicity, what do you think that we use 
\end{abstract}

\begin{CCSXML}
<ccs2012>
   <concept>       <concept_id>10003120.10003130.10003131.10003235</concept_id>
       <concept_desc>Human-centered computing~Collaborative content creation</concept_desc>
       <concept_significance>500</concept_significance>
       </concept>
 </ccs2012>
\end{CCSXML}
\ccsdesc[500]{Human-centered computing~Collaborative content creation}

\keywords{Human-AI Collaborative Writing, Transparency, AI Disclosure, Content Co-creation}

\maketitle

\section{Introduction}

With the growing adoption of AI\footnote{While AI can take various forms, in this paper, we use ``AI'' to describe technologies powered by large language models (LLMs), which specialize in processing and generating natural languages~\cite{minaee2024large}.} as a writing support, we are increasingly exposed to content shaped by AI (e.g., fully generated or lightly modified by AI)~\cite{liang2024monitoring, liang2024mapping, kobak2024delving} that blends seamlessly with human-written text, oftentimes without noticing~\cite{zhang2024llm}.
The fusion of human-written and AI-generated content raises questions about ownership, transparency, and trust of human-AI co-created content. At the core of these concerns lies writers' disclosure behavior---while disclosing contributor(s) to writing could largely improve openness and help readers calibrate trust~\cite{newresearch}, many writers have yet to adopt this practice~\cite{zhang2024secret}. 
% We aim to understand what makes a writer disclose or withhold AI's involvement in their writing process. 

Recent studies showed that writers' (non-)disclosure or authorship attribution\footnote{Although the terms are sometimes used interchangeably, we use the following operational definitions to distinguish between ``disclosure'' and ``attribution.'' 
We define ``disclosure'' as the act of indicating whether AI was involved in the writing process (e.g., a binary yes or no), whereas ``attribution'' (or ``credit assignment'') refers to more fine-grained acknowledgments (e.g., detailing how AI contributed and to what extent).} behaviors are influenced by the type of the task that a writer performs (e.g., creative writing or school assignment), perceived external judgments from readers (e.g., social bias toward AI-assisted writing)~\cite{zhang2024secret}, the type and amount of AI support (e.g., whether a writer used AI to come up with new ideas or only check for spelling and grammar errors)~\cite{he2025contributions}, writer- or AI-initiated (e.g., whether a writer requests AI support or a system initiates it)~\cite{he2025contributions}, and a writer's apprehension on perceived authenticity (e.g., uncertainty about how readers might perceive the authenticity of the written artifact)~\cite{hwang202480}. While prior works look at a subset of these factors at a time, there is a lack of a holistic investigation of how they could affect writers' disclosure decisions when other factors co-exist. 
% Mina: I commented this part out thinking that this description still feels a bit loftly and disconnected, using science of science argument rather than explicitly describing why an integrative approach is important for our set up (I couldn't understand how group synergy is relevant for our setting either, other than making a point about the integrative experiment design). 
% Taking an integrative approach to study topics where one can find many relevant factors and irreconcilable empirical results is essential, as experiments designed to test independent variables in isolation (often with different study settings) cannot be reliably integrated together~\cite{almaatouq2024beyond}\footnote{A notable example is \emph{group synergy}, where extensive empirical work has produced inconsistent results on whether group output exceeds the sum of individual performance~\cite{almaatouq2024beyond, larson2013search}}.

In this work, we synthesize findings from prior work and propose to take an integrative approach~\cite{almaatouq2024beyond} to answer the following question: \textit{What factors shape writers' decisions to disclose their AI use in writing?} 
% In social and behavioral sciences, taking an ``integrative approach'' to aggregate knowledge in a problem domain \mina{what does ``aggregate knowledge in a problem domain'' mean here?} and holistically study how various factors could together influence an outcome is deemed valuable~\cite{almaatouq2024beyond} \mina{passive voice}. \mina{I'm still finding this to be not convincing enough (it feels like a hallow statement that's not fully connected to our contest).my suggestion - move this citation to ``propose to take an integrative approach to answer the research question'', cut this sentence, and give a concrete example what potential harm there could be if no one does a holistic study}
Our work aims to offer a foundation for generating a more holistic understanding of how these elements collectively influence writers' disclosure practices.
In Section 2, we review the broader context in which this study is situated. In Section 3, we present the factors we investigate and the underlying rationales. 
\section{Background}

\paragraph{Disclosing human contributors in collaborative work} 
Collaborative work among humans existed long before AI, as do the accompanying discussions about ownership, disclosure, attribution, and credit assignment~\cite{riley1996crafting}. 
Research in behavioral science shows that people feel a sense of ownership not only over physical possessions but also over intellectual properties, and using or claiming other people's ideas without proper credit attribution is disfavored~\cite{shaw2012children, pierce2003state}. In many cases, people do not feel full ownership over co-created work~\cite{he2024ai, gero2019metaphoria}, but this does not necessarily lead to giving credit to others~\cite{draxler2024ai}. The practice of not attributing to contributors in collaborative writing is known as \emph{ghostwriting}~\cite{carvalho2021defying, gotzsche2009should, riley1996crafting}. To restrain unethical concealment of contributors, in formal contexts such as academic writing, there are established norms and community guidelines regarding who should and should not be listed as an author of a published paper~\cite{brand2015beyond}. 
In less formal environments, the standards for proper disclosure, attribution, and credit assignment are less clearly defined. In online communities, contributors oftentimes do not receive proper attribution, and the lack of economic incentives makes long-term involvement difficult to sustain~\cite{cranshaw2011polymath}. In user-generated content communities such as Scratch, 
manual credit-giving by a remixer\footnote{In Scratch, a remixer is a person who creates a new project by modifying an existing project.} to the original creator creates a sense of recognition and respect that a computer system's automatic attribution fails to replicate~\cite{monroy2011computers}. 
In addition, clear attribution to original creators or source materials is considered central to authenticity and legitimacy of creative work~\cite{sinnreich2009ethics}.

\paragraph{Disclosing AI's involvement in human-AI co-creation} 

Although many people desire to know AI's involvement in the human-AI co-creation process~\cite{avery2024attributing}, disclosing AI use has yet to become a common practice among writers~\cite{draxler2024ai}. 
Studies show that writers are far less likely to disclose or credit AI compared to human collaborators~\cite{he2024ai, draxler2024ai}. 
What drives a writer to disclose AI use, especially when disclosure is not mandated, is a complex mix of internal and external factors. These factors encourage or discourage disclosure depending on specific contexts, such as the role AI plays in the writing and how the target audience is likely to judge its use.
While several recent works studied the effects of various factors on writers' disclosure behaviors independently, we lack a holistic view of how these factors can jointly influence writers' decisions (see Appendix~\ref{sec:appendix-procedural}-\ref{sec:appendix-personal} for detailed comparison). For example, recent works found that writers tend to credit AI differently based on how they interacted with AI, including AI's contribution type~\cite{he2025contributions, rezwana2023user}, amount~\cite{he2025contributions, xu2024makes}, and initiative~\cite{he2025contributions}.
In contrast, \citet{zhang2024secret} investigated the social factors surrounding AI use in writing: Disclosing AI involvement in writing can influence how readers perceive the quality of the work~\cite{li2024does}; In turn, the perceived readers' judgments influence writers' decisions about whether to disclose~\cite{zhang2024secret}.
We systematically synthesize factors from prior work to lay the foundation for a holistic investigation.

\section{Factors that May Influence Writers' Decisions to Disclose AI Use}

Building on prior works in human-AI interaction 
% \mina{consistency - human-AI co-creation vs. human-AI interaction} 
and behavioral science, we thematically identified three groups of factors that can potentially shape writers' disclosure decisions: \emph{Procedural factors} (i.e., how writers' use of AI shape their decisions), \emph{social factors} (i.e., how others' behaviors, expectations, and judgments influence writers' decisions), and \emph{personal factors} (i.e., how individual traits, such as demographics and values, affect their decisions). 

\subsection{Procedural Factors}

Procedural factors describe how writers use AI during writing. 
These factors are drawn primarily from the collective-centered creation (CCC) theoretical framework, which conceptually discusses features of human-AI co-creation procedures to inform how credit should be distributed~\cite{khosrowi2023diffusing}.
On the other hand, some recent studies on AI-assisted writing touch on a subset of these factors and offer empirical evidence on their effects~\cite{he2025contributions, xu2024makes}. 
We incorporate insights from both sides to curate and define the following procedural factors (see Appendix~\ref{sec:appendix-procedural} for comparison to prior work).

\begin{enumerate}

    \item \textbf{Replaceability} describes how replaceable AI was to create the final written artifact. It captures the extent to which AI's contribution could be removed or substituted without significantly altering the outcome. 
    In a high-replaceability scenario, a writer can create similar content without AI, whereas in a low-replaceability scenario, a writer may not be capable of creating a comparable artifact if AI is taken away.
    % \mina{similarly confusing - it sounds like this is a test one can perform afterwards, but what we mean is really looking back in the past.}
    % we can consider the AI to have a low replaceability; in contrast, if a writer can create similar content even without AI, we can describe the AI to have a high replaceability

    \item \textbf{Effortfulness} reflects how much a writer invested in creating the written artifact when AI was involved. In some cases, a writer may put significantly less effort into writing as they used AI to automate much of the process (i.e., low effortfulness). In contrast, writers could put comparable (or even more) effort as they spent extra time on crafting good prompts or inspecting and revising AI-generated content (i.e., high effortfulness).

    \item \textbf{Intentionality} describes whether a writer intentionally steered the written artifact toward a concrete vision or specific target in their mind. In a high-intentionality scenario, a writer would use AI in a goal-oriented manner (e.g., repeatedly prompted AI to generate something they wanted), while in a low-intentionality setting, a writer might let AI take the lead (e.g., let AI generate text with minimal guidance or intervention).

    \item \textbf{Directness} depicts how closely and causally AI's contribution was represented in the final written artifact. For example, directly incorporating AI-generated text as part of a written artifact with no or minimal change could be described as ``direct'' AI use, while considering AI's feedback without copying AI-generated text into writing could be described as ``indirect'' AI use.

\end{enumerate}

\subsection{Social Factors}
%% appendix: did not look at the varying degrees of AI involvement
Social factors capture external situations that could affect writers' judgments. Humans are social animals and their behaviors are influenced by how others (may) behave or react.
For example,~\citet{zhang2024secret} found that writers' perceived judgments from readers (i.e., how writers \emph{believe} readers would think about them) lead to concealment of their use of AI. The social factors are informed by theories in behavioral science and complemented with empirical findings in HCI (see Appendix~\ref{sec:appendix-social} for comparison to prior work).

\begin{enumerate}

    \item \textbf{Detectability} refers to the probability that readers can recognize AI's contribution even without writers' explicit disclosure of AI use, be it generated entirely or modified by AI to any extent. High detectability may prompt writers to proactively disclose their AI use, while low detectability may lead them to withhold disclosure, feeling it is unlikely to be noticed or questioned.
    % , as unsuccessfully hiding AI use is unlikely to bring any benefit. 
    % To the best of our knowledge, there is no empirical study showing that detectability has a main effect on writers' disclosure so far. 
    % The interest and demand in algorithmic detectors for AI-generated content~\cite{wang2024stumbling, kushnareva2023ai} and the fact that writers care much about how readers might think about them~\cite{zhang2024secret} led us to hypothesize that detectability might be a key factor to be considered. \mina{instead of describing how we came up with this factor (move it to appendix) and reiterating background work, describe how high vs. low detectability might affect writers' behavior instead} 

    \item \textbf{Regulatory penalty for misclaim} reflects the punishment for misclaiming any AI's contribution to the written artifact (including not disclosing AI use when AI was involved in writing). When strong regulatory penalties are in place, violating disclosure rules can result in consequences that were directly and explicitly stated in the rules~\cite{acmpolicy}. 
    % According to findings from participatory workshops about transparent AI disclosure, people are most concerned with ``legal compliance and AI disclosure'' and ``policy and regulatory impact''~\cite{el2024transparent}. 
    Notably, regulatory penalty does not always lead to disclosure~\cite{hacker2023regulating}---some writers may intentionally hide their AI use (e.g., when they believe their AI use cannot be detected without disclosure).

    \item \textbf{Perceived negative effects of disclosure} refers to potential negative consequences associated with disclosing AI use in writing. Examples include social bias and stigma toward AI-generated content or writers who use AI. Compared to regulatory penalty for misclaim, these potential consequences are implicit and unlikely to pose direct and immediate negative consequences to the writers. 
    Some examples of negative consequences could be the decrease in perceived quality~\cite{li2024does} and sincerity~\cite{zhang2024secret} of the written artifact.

    \item \textbf{Perceived positive effects of disclosure} refers to potential benefits that disclosing AI use could bring. For example, if a writer discloses their AI use, readers may perceive the writer to be honest as they avoid unethically taking credit for AI's contribution. As disclosure can promote transparency and foster trust~\cite{renieris2024artificial, liao2023ai}, people may disclose their AI use for the purpose of impression management~\cite{tedeschi2013impression}. 
    % \mina{the two parts in the last sentence felt a bit disconnected, so I connected them but I'm not sure if this is what you intended; what do you think?}

    \item \textbf{Prevalence of disclosure} reflects how often others disclose their AI use. 
    % Informed by the social learning theory~\cite{pratt2010empirical} and the effects of social norms~\cite{cialdini1998social}, \mina{describe these two to be self-contained; move to appendix for consistency}
    Prevalent disclosure of AI use may encourage more writers to follow the social norm~\cite{cialdini1998social, pratt2010empirical}, whereas low prevalence may make writers think that disclosure is unnecessary.

\end{enumerate}

\subsection{Personal Factors}

Personal factors reflect writers' individual differences in background, preferences, and values. Prior work showed mixed results: \citet{zhang2024secret} observed that writers' backgrounds and personality traits did not significantly impact their tendency to hide AI use~\cite{zhang2024secret}, whereas \citet{he2025contributions} found that some writers develop ``personal rubrics'' that inform how to credit AI's contribution, which revealed that individual differences may affect writers' disclosure decisions. Thus, we decided to include personal factors in our work for further investigation. 
Personal factors listed below are mostly informed by~\citet{zhang2024secret} (see Appendix~\ref{sec:appendix-personal} for comparison to prior work).

% \mina{based on our discussion, what's your plan for these factors? if we were to talk about various types of self-efficacy under one factor, I think we need to rewrite the description so that it's not like we describe one version and then add a new hypothesis in the end. instead, we could first say that there can be two types of self-efficacy and then talk about them. same for internal judgments.}

\begin{enumerate}
    \item \textbf{Self-efficacy} measures people's beliefs about their capability to complete a task successfully~\cite{bandura1997self}. Two types of self-efficacy are relevant to AI-assisted writing: Self-efficacy of writing itself and self-efficacy of using AI for writing. When writers feel confident about their ability in writing or using AI to support writing, they may consider concealing AI use to be unnecessary~\cite{zhang2024secret}. 
    % \mina{this paper also talks about self-efficacy; take a look and see if we want to cite it: “If the Machine Is As Good As Me, Then What Use Am I?” – How the Use of ChatGPT Changes Young Professionals’ Perception of Productivity and Accomplishment}
    % \citet{zhang2024secret} showed evidence that low self-efficacy of writing may cause writers to conceal their use of AI. Low self-efficacy in using AI to assist with writing may cause a similar effect.  
    % Two dimensions are relevant to this study's context: self-efficacy of writing and self-efficacy of using AI to support writing.  \mina{explain why we are merging the two into one factor?}
    
    \item \textbf{Internal judgments} captures how a writer reflects on or assesses their own behaviors. Two types of internal judgments are relevant to AI-assisted writing: Internal judgments on using AI (e.g., whether using AI in writing is moral and ethical) and internal judgments on disclosure (e.g., how much a writer values transparency in communication). A writer who thinks using AI in writing is unethical may be less likely to disclose their AI use, while a writer who values transparency may be more likely to disclose. 
    % ~\citet{zhang2024secret} considered that internal judgments on using AI (e.g., whether using AI in writing is moral and ethical) affects writers' disclosure. In addition, we hypothesize that the internal judgments on the disclosing behavior itself (e.g., how much a writer values transparency in communication) might also influence their disclosure behavior. 
    
    \item \textbf{Demographics} (e.g., gender and age) have been found to influence people's use of AI~\cite{draxler2023gender} and intentions of performing secrecy or (un)ethical behaviors in certain contexts~\cite{zhang2024secret, leonard2004influences, dawson1997ethical}. 
    % \mina{can you elaborate on this? it seems a bit jarring to see a generic statement that indicates certain gender and age groups perform unethical behaviors...}. 
    Since the disclosure of AI use is tied to ethical considerations, demographic background might be a potential factor that affects writers' disclosure decisions. 
    % \mina{I'm not sure what the last sentence is implying... cut?}
    
    % As disclosing contributor(s) to a writing artifact is inherently related to ethics, demographic differences may also be a factor worth investigating. 
    
    % \item \textbf{Demographics} have been found to influence people's intentions of performing (un)ethical behaviors in certain contexts~\cite{zhang2024secret, leonard2004influences, dawson1997ethical}. As disclosing contributor(s) to a writing artifact is inherently related to ethics, demographic differences may also be a factor worth investigating. 
\end{enumerate}

\subsection{Future work}
% \mina{I think the current version is more of a conclusion than actual discussion...
% what do we want to convey in this paragraph?}

Although this list may not cover every possible factor that shapes writers' decisions to disclose AI use, our aim is to highlight key considerations informed by prior works.  
When a writer thinks about whether to disclose AI's involvement in writing, they might face a situation jointly shaped by these factors. 
While existing studies suggest that each factor may independently influence writers' disclosure decisions, less is known about how these factors work together. 
For the next step, we plan to take an integrative approach to explore these factors' combined effects and measure their relative significance in influencing writers' disclosure decisions. 
% In the real-world scenario, when writers decide whether to disclose AI's involvement in the writing, they often find themselves navigating through the complex dynamics jointly shaped by an array of factors, each may drag them into slightly different directions. 

% \mina{how about mentioning 1) these factors can characterize and describe a task (use of proposed factors), 2) future research can measure the relative importance of these factors on decision-making (follow-up study with the factors), and one or two other things to demonstrate the use/value/limitations of the proposed factors?}

% \subsection{Study Overview}
% \subsection{Envisioned Contribution and Broader Impact}

% broad coverage of relevant factors that more realistically reflect writers' situations (compare to prev studies, mention beyond playing with 20 questions paper)
% account for changing society norms, informing possible future changes
% identify potential misalignment in writers' practices and readers' expectations, offer insights into design interventions that encourage transparent AI use to close the gap

% \subsection{Data analysis}

\bibliographystyle{ACM-Reference-Format}
\bibliography{main}

%%% -*-BibTeX-*-
%%% Do NOT edit. File created by BibTeX with style
%%% ACM-Reference-Format-Journals [18-Jan-2012].

\begin{thebibliography}{39}

%%% ====================================================================
%%% NOTE TO THE USER: you can override these defaults by providing
%%% customized versions of any of these macros before the \bibliography
%%% command.  Each of them MUST provide its own final punctuation,
%%% except for \shownote{}, \showDOI{}, and \showURL{}.  The latter two
%%% do not use final punctuation, in order to avoid confusing it with
%%% the Web address.
%%%
%%% To suppress output of a particular field, define its macro to expand
%%% to an empty string, or better, \unskip, like this:
%%%
%%% \newcommand{\showDOI}[1]{\unskip}   % LaTeX syntax
%%%
%%% \def \showDOI #1{\unskip}           % plain TeX syntax
%%%
%%% ====================================================================

\ifx \showCODEN    \undefined \def \showCODEN     #1{\unskip}     \fi
\ifx \showDOI      \undefined \def \showDOI       #1{#1}\fi
\ifx \showISBNx    \undefined \def \showISBNx     #1{\unskip}     \fi
\ifx \showISBNxiii \undefined \def \showISBNxiii  #1{\unskip}     \fi
\ifx \showISSN     \undefined \def \showISSN      #1{\unskip}     \fi
\ifx \showLCCN     \undefined \def \showLCCN      #1{\unskip}     \fi
\ifx \shownote     \undefined \def \shownote      #1{#1}          \fi
\ifx \showarticletitle \undefined \def \showarticletitle #1{#1}   \fi
\ifx \showURL      \undefined \def \showURL       {\relax}        \fi
% The following commands are used for tagged output and should be
% invisible to TeX
\providecommand\bibfield[2]{#2}
\providecommand\bibinfo[2]{#2}
\providecommand\natexlab[1]{#1}
\providecommand\showeprint[2][]{arXiv:#2}

\bibitem[Almaatouq et~al\mbox{.}(2024)]%
        {almaatouq2024beyond}
\bibfield{author}{\bibinfo{person}{Abdullah Almaatouq}, \bibinfo{person}{Thomas~L Griffiths}, \bibinfo{person}{Jordan~W Suchow}, \bibinfo{person}{Mark~E Whiting}, \bibinfo{person}{James Evans}, {and} \bibinfo{person}{Duncan~J Watts}.} \bibinfo{year}{2024}\natexlab{}.
\newblock \showarticletitle{Beyond playing 20 questions with nature: Integrative experiment design in the social and behavioral sciences}.
\newblock \bibinfo{journal}{\emph{Behavioral and Brain Sciences}}  \bibinfo{volume}{47} (\bibinfo{year}{2024}), \bibinfo{pages}{e33}.
\newblock


\bibitem[Avery et~al\mbox{.}(2024)]%
        {avery2024attributing}
\bibfield{author}{\bibinfo{person}{Joseph~J Avery}, \bibinfo{person}{Patricia~S{\'a}nchez Abril}, {and} \bibinfo{person}{Alissa del Riego}.} \bibinfo{year}{2024}\natexlab{}.
\newblock \showarticletitle{Attributing AI Authorship: Towards a System of Icons for Legal and Ethical Disclosure}.
\newblock \bibinfo{journal}{\emph{Nw. J. Tech. \& Intell. Prop.}}  \bibinfo{volume}{22} (\bibinfo{year}{2024}), \bibinfo{pages}{1}.
\newblock


\bibitem[Bandura and Wessels(1997)]%
        {bandura1997self}
\bibfield{author}{\bibinfo{person}{Albert Bandura} {and} \bibinfo{person}{Sebastian Wessels}.} \bibinfo{year}{1997}\natexlab{}.
\newblock \bibinfo{booktitle}{\emph{Self-efficacy}}.
\newblock \bibinfo{publisher}{Cambridge University Press Cambridge}.
\newblock


\bibitem[Board(nd)]%
        {acmpolicy}
\bibfield{author}{\bibinfo{person}{ACM~Publications Board}.} \bibinfo{year}{n.d.}\natexlab{}.
\newblock \bibinfo{title}{ACM Policy on Authorship}.
\newblock \bibinfo{howpublished}{\url{https://www.acm.org/publications/policies/new-acm-policy-on-authorship}}.
\newblock
\newblock
\shownote{Accessed: 2025-04-15}.


\bibitem[Brand et~al\mbox{.}(2015)]%
        {brand2015beyond}
\bibfield{author}{\bibinfo{person}{Amy Brand}, \bibinfo{person}{Liz Allen}, \bibinfo{person}{Micah Altman}, \bibinfo{person}{Marjorie Hlava}, {and} \bibinfo{person}{Jo Scott}.} \bibinfo{year}{2015}\natexlab{}.
\newblock \showarticletitle{Beyond authorship: Attribution, contribution, collaboration, and credit.}
\newblock \bibinfo{journal}{\emph{Learned Publishing}} \bibinfo{volume}{28}, \bibinfo{number}{2} (\bibinfo{year}{2015}).
\newblock


\bibitem[Carvalho et~al\mbox{.}(2021)]%
        {carvalho2021defying}
\bibfield{author}{\bibinfo{person}{John Carvalho}, \bibinfo{person}{Angie Chung}, {and} \bibinfo{person}{Michael Koliska}.} \bibinfo{year}{2021}\natexlab{}.
\newblock \showarticletitle{Defying transparency: Ghostwriting from the Jazz Age to social media}.
\newblock \bibinfo{journal}{\emph{Journalism}} \bibinfo{volume}{22}, \bibinfo{number}{3} (\bibinfo{year}{2021}), \bibinfo{pages}{709--725}.
\newblock


\bibitem[Cialdini and Trost(1998)]%
        {cialdini1998social}
\bibfield{author}{\bibinfo{person}{Robert~B Cialdini} {and} \bibinfo{person}{Melanie~R Trost}.} \bibinfo{year}{1998}\natexlab{}.
\newblock \showarticletitle{Social influence: Social norms, conformity and compliance.}
\newblock  (\bibinfo{year}{1998}).
\newblock


\bibitem[Cranshaw and Kittur(2011)]%
        {cranshaw2011polymath}
\bibfield{author}{\bibinfo{person}{Justin Cranshaw} {and} \bibinfo{person}{Aniket Kittur}.} \bibinfo{year}{2011}\natexlab{}.
\newblock \showarticletitle{The polymath project: lessons from a successful online collaboration in mathematics}. In \bibinfo{booktitle}{\emph{Proceedings of the SIGCHI conference on human factors in computing systems}}. \bibinfo{pages}{1865--1874}.
\newblock


\bibitem[Dawson(1997)]%
        {dawson1997ethical}
\bibfield{author}{\bibinfo{person}{Leslie~M Dawson}.} \bibinfo{year}{1997}\natexlab{}.
\newblock \showarticletitle{Ethical differences between men and women in the sales profession}.
\newblock \bibinfo{journal}{\emph{Journal of Business Ethics}}  \bibinfo{volume}{16} (\bibinfo{year}{1997}), \bibinfo{pages}{1143--1152}.
\newblock


\bibitem[Draxler et~al\mbox{.}(2023)]%
        {draxler2023gender}
\bibfield{author}{\bibinfo{person}{Fiona Draxler}, \bibinfo{person}{Daniel Buschek}, \bibinfo{person}{Mikke Tavast}, \bibinfo{person}{Perttu H{\"a}m{\"a}l{\"a}inen}, \bibinfo{person}{Albrecht Schmidt}, \bibinfo{person}{Juhi Kulshrestha}, {and} \bibinfo{person}{Robin Welsch}.} \bibinfo{year}{2023}\natexlab{}.
\newblock \showarticletitle{Gender, age, and technology education influence the adoption and appropriation of LLMs}.
\newblock \bibinfo{journal}{\emph{arXiv preprint arXiv:2310.06556}} (\bibinfo{year}{2023}).
\newblock


\bibitem[Draxler et~al\mbox{.}(2024)]%
        {draxler2024ai}
\bibfield{author}{\bibinfo{person}{Fiona Draxler}, \bibinfo{person}{Anna Werner}, \bibinfo{person}{Florian Lehmann}, \bibinfo{person}{Matthias Hoppe}, \bibinfo{person}{Albrecht Schmidt}, \bibinfo{person}{Daniel Buschek}, {and} \bibinfo{person}{Robin Welsch}.} \bibinfo{year}{2024}\natexlab{}.
\newblock \showarticletitle{The AI ghostwriter effect: When users do not perceive ownership of AI-generated text but self-declare as authors}.
\newblock \bibinfo{journal}{\emph{ACM Transactions on Computer-Human Interaction}} \bibinfo{volume}{31}, \bibinfo{number}{2} (\bibinfo{year}{2024}), \bibinfo{pages}{1--40}.
\newblock


\bibitem[Gero and Chilton(2019)]%
        {gero2019metaphoria}
\bibfield{author}{\bibinfo{person}{Katy~Ilonka Gero} {and} \bibinfo{person}{Lydia~B Chilton}.} \bibinfo{year}{2019}\natexlab{}.
\newblock \showarticletitle{Metaphoria: An algorithmic companion for metaphor creation}. In \bibinfo{booktitle}{\emph{Proceedings of the 2019 CHI conference on human factors in computing systems}}. \bibinfo{pages}{1--12}.
\newblock


\bibitem[G{\o}tzsche et~al\mbox{.}(2009)]%
        {gotzsche2009should}
\bibfield{author}{\bibinfo{person}{Peter~C G{\o}tzsche}, \bibinfo{person}{Jerome~P Kassirer}, \bibinfo{person}{Karen~L Woolley}, \bibinfo{person}{Elizabeth Wager}, \bibinfo{person}{Adam Jacobs}, \bibinfo{person}{Art Gertel}, {and} \bibinfo{person}{Cindy Hamilton}.} \bibinfo{year}{2009}\natexlab{}.
\newblock \showarticletitle{What should be done to tackle ghostwriting in the medical literature?}
\newblock \bibinfo{journal}{\emph{PLoS Medicine}} \bibinfo{volume}{6}, \bibinfo{number}{2} (\bibinfo{year}{2009}), \bibinfo{pages}{e1000023}.
\newblock


\bibitem[Hacker et~al\mbox{.}(2023)]%
        {hacker2023regulating}
\bibfield{author}{\bibinfo{person}{Philipp Hacker}, \bibinfo{person}{Andreas Engel}, {and} \bibinfo{person}{Marco Mauer}.} \bibinfo{year}{2023}\natexlab{}.
\newblock \showarticletitle{Regulating ChatGPT and other large generative AI models}. In \bibinfo{booktitle}{\emph{Proceedings of the 2023 ACM conference on fairness, accountability, and transparency}}. \bibinfo{pages}{1112--1123}.
\newblock


\bibitem[He et~al\mbox{.}(2024)]%
        {he2024ai}
\bibfield{author}{\bibinfo{person}{Jessica He}, \bibinfo{person}{Stephanie Houde}, \bibinfo{person}{Gabriel~E Gonzalez}, \bibinfo{person}{Dar{\'\i}o~Andr{\'e}s Silva~Moran}, \bibinfo{person}{Steven~I Ross}, \bibinfo{person}{Michael Muller}, {and} \bibinfo{person}{Justin~D Weisz}.} \bibinfo{year}{2024}\natexlab{}.
\newblock \showarticletitle{AI and the Future of Collaborative Work: Group Ideation with an LLM in a Virtual Canvas}. In \bibinfo{booktitle}{\emph{Proceedings of the 3rd Annual Meeting of the Symposium on Human-Computer Interaction for Work}}. \bibinfo{pages}{1--14}.
\newblock


\bibitem[He et~al\mbox{.}(2025)]%
        {he2025contributions}
\bibfield{author}{\bibinfo{person}{Jessica He}, \bibinfo{person}{Stephanie Houde}, {and} \bibinfo{person}{Justin~D Weisz}.} \bibinfo{year}{2025}\natexlab{}.
\newblock \showarticletitle{Which Contributions Deserve Credit? Perceptions of Attribution in Human-AI Co-Creation}.
\newblock \bibinfo{journal}{\emph{arXiv preprint arXiv:2502.18357}} (\bibinfo{year}{2025}).
\newblock


\bibitem[Hwang et~al\mbox{.}(2024)]%
        {hwang202480}
\bibfield{author}{\bibinfo{person}{Angel Hsing-Chi Hwang}, \bibinfo{person}{Q~Vera Liao}, \bibinfo{person}{Su~Lin Blodgett}, \bibinfo{person}{Alexandra Olteanu}, {and} \bibinfo{person}{Adam Trischler}.} \bibinfo{year}{2024}\natexlab{}.
\newblock \showarticletitle{" It was 80\% me, 20\% AI": Seeking Authenticity in Co-Writing with Large Language Models}.
\newblock \bibinfo{journal}{\emph{arXiv preprint arXiv:2411.13032}} (\bibinfo{year}{2024}).
\newblock


\bibitem[Khosrowi et~al\mbox{.}(2023)]%
        {khosrowi2023diffusing}
\bibfield{author}{\bibinfo{person}{Donal Khosrowi}, \bibinfo{person}{Finola Finn}, {and} \bibinfo{person}{Elinor Clark}.} \bibinfo{year}{2023}\natexlab{}.
\newblock \showarticletitle{Diffusing the creator: Attributing credit for generative AI outputs}. In \bibinfo{booktitle}{\emph{Proceedings of the 2023 AAAI/ACM Conference on AI, Ethics, and Society}}. \bibinfo{pages}{890--900}.
\newblock


\bibitem[Kobak et~al\mbox{.}(2024)]%
        {kobak2024delving}
\bibfield{author}{\bibinfo{person}{Dmitry Kobak}, \bibinfo{person}{Rita Gonz{\'a}lez-M{\'a}rquez}, \bibinfo{person}{Em{\H{o}}ke-{\'A}gnes Horv{\'a}t}, {and} \bibinfo{person}{Jan Lause}.} \bibinfo{year}{2024}\natexlab{}.
\newblock \showarticletitle{Delving into ChatGPT usage in academic writing through excess vocabulary}.
\newblock \bibinfo{journal}{\emph{arXiv preprint arXiv:2406.07016}} (\bibinfo{year}{2024}).
\newblock


\bibitem[Leonard et~al\mbox{.}(2004)]%
        {leonard2004influences}
\bibfield{author}{\bibinfo{person}{Lori~NK Leonard}, \bibinfo{person}{Timothy~Paul Cronan}, {and} \bibinfo{person}{Jennifer Kreie}.} \bibinfo{year}{2004}\natexlab{}.
\newblock \showarticletitle{What influences IT ethical behavior intentions—planned behavior, reasoned action, perceived importance, or individual characteristics?}
\newblock \bibinfo{journal}{\emph{Information \& Management}} \bibinfo{volume}{42}, \bibinfo{number}{1} (\bibinfo{year}{2004}), \bibinfo{pages}{143--158}.
\newblock


\bibitem[Li et~al\mbox{.}(2024)]%
        {li2024does}
\bibfield{author}{\bibinfo{person}{Zhuoyan Li}, \bibinfo{person}{Chen Liang}, \bibinfo{person}{Jing Peng}, {and} \bibinfo{person}{Ming Yin}.} \bibinfo{year}{2024}\natexlab{}.
\newblock \showarticletitle{How Does the Disclosure of AI Assistance Affect the Perceptions of Writing?}
\newblock \bibinfo{journal}{\emph{arXiv preprint arXiv:2410.04545}} (\bibinfo{year}{2024}).
\newblock


\bibitem[Liang et~al\mbox{.}(2024a)]%
        {liang2024monitoring}
\bibfield{author}{\bibinfo{person}{Weixin Liang}, \bibinfo{person}{Zachary Izzo}, \bibinfo{person}{Yaohui Zhang}, \bibinfo{person}{Haley Lepp}, \bibinfo{person}{Hancheng Cao}, \bibinfo{person}{Xuandong Zhao}, \bibinfo{person}{Lingjiao Chen}, \bibinfo{person}{Haotian Ye}, \bibinfo{person}{Sheng Liu}, \bibinfo{person}{Zhi Huang}, {et~al\mbox{.}}} \bibinfo{year}{2024}\natexlab{a}.
\newblock \showarticletitle{Monitoring ai-modified content at scale: A case study on the impact of chatgpt on ai conference peer reviews}.
\newblock \bibinfo{journal}{\emph{arXiv preprint arXiv:2403.07183}} (\bibinfo{year}{2024}).
\newblock


\bibitem[Liang et~al\mbox{.}(2024b)]%
        {liang2024mapping}
\bibfield{author}{\bibinfo{person}{Weixin Liang}, \bibinfo{person}{Yaohui Zhang}, \bibinfo{person}{Zhengxuan Wu}, \bibinfo{person}{Haley Lepp}, \bibinfo{person}{Wenlong Ji}, \bibinfo{person}{Xuandong Zhao}, \bibinfo{person}{Hancheng Cao}, \bibinfo{person}{Sheng Liu}, \bibinfo{person}{Siyu He}, \bibinfo{person}{Zhi Huang}, {et~al\mbox{.}}} \bibinfo{year}{2024}\natexlab{b}.
\newblock \showarticletitle{Mapping the increasing use of LLMs in scientific papers}.
\newblock \bibinfo{journal}{\emph{arXiv preprint arXiv:2404.01268}} (\bibinfo{year}{2024}).
\newblock


\bibitem[Liao and Vaughan(2023)]%
        {liao2023ai}
\bibfield{author}{\bibinfo{person}{Q~Vera Liao} {and} \bibinfo{person}{Jennifer~Wortman Vaughan}.} \bibinfo{year}{2023}\natexlab{}.
\newblock \showarticletitle{Ai transparency in the age of llms: A human-centered research roadmap}.
\newblock \bibinfo{journal}{\emph{arXiv preprint arXiv:2306.01941}}  \bibinfo{volume}{10} (\bibinfo{year}{2023}).
\newblock


\bibitem[Minaee et~al\mbox{.}(2024)]%
        {minaee2024large}
\bibfield{author}{\bibinfo{person}{Shervin Minaee}, \bibinfo{person}{Tomas Mikolov}, \bibinfo{person}{Narjes Nikzad}, \bibinfo{person}{Meysam Chenaghlu}, \bibinfo{person}{Richard Socher}, \bibinfo{person}{Xavier Amatriain}, {and} \bibinfo{person}{Jianfeng Gao}.} \bibinfo{year}{2024}\natexlab{}.
\newblock \showarticletitle{Large language models: A survey, 2024}.
\newblock \bibinfo{journal}{\emph{arXiv preprint arXiv:2402.06196}} (\bibinfo{year}{2024}).
\newblock


\bibitem[Monroy-Hern{\'a}ndez et~al\mbox{.}(2011)]%
        {monroy2011computers}
\bibfield{author}{\bibinfo{person}{Andr{\'e}s Monroy-Hern{\'a}ndez}, \bibinfo{person}{Benjamin~Mako Hill}, \bibinfo{person}{Jazmin Gonzalez-Rivero}, {and} \bibinfo{person}{Danah Boyd}.} \bibinfo{year}{2011}\natexlab{}.
\newblock \showarticletitle{Computers can't give credit: How automatic attribution falls short in an online remixing community}. In \bibinfo{booktitle}{\emph{Proceedings of the SIGCHI Conference on Human Factors in Computing Systems}}. \bibinfo{pages}{3421--3430}.
\newblock


\bibitem[Pierce et~al\mbox{.}(2003)]%
        {pierce2003state}
\bibfield{author}{\bibinfo{person}{Jon~L Pierce}, \bibinfo{person}{Tatiana Kostova}, {and} \bibinfo{person}{Kurt~T Dirks}.} \bibinfo{year}{2003}\natexlab{}.
\newblock \showarticletitle{The state of psychological ownership: Integrating and extending a century of research}.
\newblock \bibinfo{journal}{\emph{Review of general psychology}} \bibinfo{volume}{7}, \bibinfo{number}{1} (\bibinfo{year}{2003}), \bibinfo{pages}{84--107}.
\newblock


\bibitem[Pratt et~al\mbox{.}(2010)]%
        {pratt2010empirical}
\bibfield{author}{\bibinfo{person}{Travis~C Pratt}, \bibinfo{person}{Francis~T Cullen}, \bibinfo{person}{Christine~S Sellers}, \bibinfo{person}{L Thomas Winfree~Jr}, \bibinfo{person}{Tamara~D Madensen}, \bibinfo{person}{Leah~E Daigle}, \bibinfo{person}{Noelle~E Fearn}, {and} \bibinfo{person}{Jacinta~M Gau}.} \bibinfo{year}{2010}\natexlab{}.
\newblock \showarticletitle{The empirical status of social learning theory: A meta-analysis}.
\newblock \bibinfo{journal}{\emph{Justice Quarterly}} \bibinfo{volume}{27}, \bibinfo{number}{6} (\bibinfo{year}{2010}), \bibinfo{pages}{765--802}.
\newblock


\bibitem[Renieris et~al\mbox{.}(2024)]%
        {renieris2024artificial}
\bibfield{author}{\bibinfo{person}{Elizabeth~M Renieris}, \bibinfo{person}{David Kiron}, {and} \bibinfo{person}{Steven Mills}.} \bibinfo{year}{2024}\natexlab{}.
\newblock \showarticletitle{Artificial Intelligence Disclosures Are Key to Customer Trust}.
\newblock \bibinfo{journal}{\emph{MIT Sloan Management Review (Online)}} (\bibinfo{year}{2024}), \bibinfo{pages}{1--4}.
\newblock


\bibitem[Rezwana and Maher(2023)]%
        {rezwana2023user}
\bibfield{author}{\bibinfo{person}{Jeba Rezwana} {and} \bibinfo{person}{Mary~Lou Maher}.} \bibinfo{year}{2023}\natexlab{}.
\newblock \showarticletitle{User perspectives on ethical challenges in human-AI co-creativity: A design fiction study}. In \bibinfo{booktitle}{\emph{Proceedings of the 15th Conference on Creativity and Cognition}}. \bibinfo{pages}{62--74}.
\newblock


\bibitem[Riley and Brown(1996)]%
        {riley1996crafting}
\bibfield{author}{\bibinfo{person}{Linda~A Riley} {and} \bibinfo{person}{Stuart~C Brown}.} \bibinfo{year}{1996}\natexlab{}.
\newblock \showarticletitle{Crafting a public image: An empirical study of the ethics of ghostwriting}.
\newblock \bibinfo{journal}{\emph{Journal of Business Ethics}}  \bibinfo{volume}{15} (\bibinfo{year}{1996}), \bibinfo{pages}{711--720}.
\newblock


\bibitem[Shaw et~al\mbox{.}(2012)]%
        {shaw2012children}
\bibfield{author}{\bibinfo{person}{Alex Shaw}, \bibinfo{person}{Vivian Li}, {and} \bibinfo{person}{Kristina~R Olson}.} \bibinfo{year}{2012}\natexlab{}.
\newblock \showarticletitle{Children apply principles of physical ownership to ideas}.
\newblock \bibinfo{journal}{\emph{Cognitive science}} \bibinfo{volume}{36}, \bibinfo{number}{8} (\bibinfo{year}{2012}), \bibinfo{pages}{1383--1403}.
\newblock


\bibitem[Sinnreich et~al\mbox{.}(2009)]%
        {sinnreich2009ethics}
\bibfield{author}{\bibinfo{person}{Aram Sinnreich}, \bibinfo{person}{Mark Latonero}, {and} \bibinfo{person}{Marissa Gluck}.} \bibinfo{year}{2009}\natexlab{}.
\newblock \showarticletitle{Ethics reconfigured: How today's media consumers evaluate the role of creative reappropriation}.
\newblock \bibinfo{journal}{\emph{Information, Communication \& Society}} \bibinfo{volume}{12}, \bibinfo{number}{8} (\bibinfo{year}{2009}), \bibinfo{pages}{1242--1260}.
\newblock


\bibitem[Tedeschi(2013)]%
        {tedeschi2013impression}
\bibfield{author}{\bibinfo{person}{James~T Tedeschi}.} \bibinfo{year}{2013}\natexlab{}.
\newblock \bibinfo{booktitle}{\emph{Impression management theory and social psychological research}}.
\newblock \bibinfo{publisher}{Academic Press}.
\newblock


\bibitem[Walsh(2024)]%
        {newresearch}
\bibfield{author}{\bibinfo{person}{Lynn Walsh}.} \bibinfo{year}{2024}\natexlab{}.
\newblock \bibinfo{title}{New research: Journalists should disclose their use of AI. Here's how.}
\newblock \bibinfo{howpublished}{\url{https://trustingnews.org/trusting-news-artificial-intelligence-ai-research-newsroom-cohort/}}.
\newblock
\newblock
\shownote{Accessed: 2025-05-25}.


\bibitem[Wan et~al\mbox{.}(2024)]%
        {wan2024coco}
\bibfield{author}{\bibinfo{person}{Ruyuan Wan}, \bibinfo{person}{Simret~Araya Gebreegziabher}, \bibinfo{person}{Toby Jia-Jun Li}, {and} \bibinfo{person}{Karla Badillo-Urquiola}.} \bibinfo{year}{2024}\natexlab{}.
\newblock \showarticletitle{CoCo Matrix: Taxonomy of Cognitive Contributions in Co-writing with Intelligent Agents}. In \bibinfo{booktitle}{\emph{Proceedings of the 16th Conference on Creativity \& Cognition}}. \bibinfo{pages}{504--511}.
\newblock


\bibitem[Xu et~al\mbox{.}(2024)]%
        {xu2024makes}
\bibfield{author}{\bibinfo{person}{Yuxin Xu}, \bibinfo{person}{Mengqiu Cheng}, {and} \bibinfo{person}{Anastasia Kuzminykh}.} \bibinfo{year}{2024}\natexlab{}.
\newblock \showarticletitle{What Makes It Mine? Exploring Psychological Ownership over Human-AI Co-Creations}. In \bibinfo{booktitle}{\emph{Proceedings of the 50th Graphics Interface Conference}}. \bibinfo{pages}{1--8}.
\newblock


\bibitem[Zhang et~al\mbox{.}(2024a)]%
        {zhang2024llm}
\bibfield{author}{\bibinfo{person}{Qihui Zhang}, \bibinfo{person}{Chujie Gao}, \bibinfo{person}{Dongping Chen}, \bibinfo{person}{Yue Huang}, \bibinfo{person}{Yixin Huang}, \bibinfo{person}{Zhenyang Sun}, \bibinfo{person}{Shilin Zhang}, \bibinfo{person}{Weiye Li}, \bibinfo{person}{Zhengyan Fu}, \bibinfo{person}{Yao Wan}, {et~al\mbox{.}}} \bibinfo{year}{2024}\natexlab{a}.
\newblock \showarticletitle{LLM-as-a-coauthor: Can mixed human-written and machine-generated text be detected?}
\newblock \bibinfo{journal}{\emph{arXiv preprint arXiv:2401.05952}} (\bibinfo{year}{2024}).
\newblock


\bibitem[Zhang et~al\mbox{.}(2024b)]%
        {zhang2024secret}
\bibfield{author}{\bibinfo{person}{Zhiping Zhang}, \bibinfo{person}{Chenxinran Shen}, \bibinfo{person}{Bingsheng Yao}, \bibinfo{person}{Dakuo Wang}, {and} \bibinfo{person}{Tianshi Li}.} \bibinfo{year}{2024}\natexlab{b}.
\newblock \showarticletitle{Secret Use of Large Language Model (LLM)}.
\newblock \bibinfo{journal}{\emph{arXiv preprint arXiv:2409.19450}} (\bibinfo{year}{2024}).
\newblock


\end{thebibliography}

\appendix
\section{Comparison to Prior Work}
\subsection{Procedural Factors}
\label{sec:appendix-procedural}

Since each procedural factor we described has been mentioned in more than one prior work with different names, in Table~\ref{table:1}, we list the sources of the procedural factors along with their original names.

% \begin{table}[ht]

% \small
% \centering
% \begin{tabular}{ p{8em} p{35em}} 
% \textbf{\textit{Procedural factors}} & \textbf{\textit{Sources}}\\
% \toprule
% Replaceability
% & Derived from the CCC framework's ``relevance/(non-)redundancy and control''~\cite{khosrowi2023diffusing}. Other works referred to similar concepts as ``contribution type''~\cite{he2025contributions}, ``AI ability ''~\cite{rezwana2023user}, and ``originality of co-creation''~\cite{xu2024makes}.
% \\
% \midrule

% Effortfulness
% & Derived from the ``time/effort'' dimension in the CCC framework~\cite{khosrowi2023diffusing}. Other works capture it with terms such as ``amount of effort''~\cite{xu2024makes} and its flip side as (AI's) ``contribution amount''~\cite{he2025contributions}.\\
% \midrule
% Intentionality
% & Discussed as ``leadership''~\cite{khosrowi2023diffusing, rezwana2023user}, ``sense of control in the process''~\cite{xu2024makes}, and ``entropy''~\cite{wan2024coco}.\\
% \midrule

% Directness
% & Originated from the CCC framework's Directness~\cite{khosrowi2023diffusing}. Strongly related to ``initiative'' (capturing whether AI contributed actual content of the writing)~\cite {he2025contributions} and ``information gain'' (where AI's direct contribution can be mapped to the writer's high information gain)~\cite{wan2024coco}.\\  
% \bottomrule
% \end{tabular}
% \caption{Mapping procedural factors to sources}
% \label{table:1}
% \end{table}

\begin{table}[h!]
\centering
\renewcommand\theadfont{\bfseries}
\begin{tabular}{@{}>{\raggedright\arraybackslash}p{2.2cm}   % Procedural factors
                >{\raggedright\arraybackslash}p{2.2cm}     % Khosrowi
                >{\raggedright\arraybackslash}p{2.2cm}     % He et al.
                >{\raggedright\arraybackslash}p{2.2cm}     % Rezwana and Maher
                >{\raggedright\arraybackslash}p{2.2cm}     % Xu et al.
                >{\raggedright\arraybackslash}p{2.1cm}@{}} % Wan et al.

\textbf{Factors} & \multicolumn{5}{c}{\textbf{Sources}} \\
\addlinespace
& \textit{\citet{khosrowi2023diffusing}} & \emph{\citet{he2025contributions}} & \emph{\citet{rezwana2023user}} & \emph{\citet{xu2024makes}} & \emph{\citet{wan2024coco}} \\
\addlinespace
\hline
\textbf{Replaceability} & Relevance/\makecell{(non-)redundancy} and control & Contribution type & AI ability & Originality of co-creation & \\
\hline
\textbf{Effortfulness} & Time/effort & Contribution amount & & Amount of effort &  \\
\hline
\textbf{Intentionality} & Leadership & & Leadership & Control in process & Entropy \\
\hline
\textbf{Directness} & Directness & Initiative & & & Information gain\\
\hline
\end{tabular}
\caption{Mapping procedural factors to sources}
\label{table:1}
\end{table}

\subsection{Social Factors}
\label{sec:appendix-social}

Very few works studied the impact of social factors on disclosing AI use (in the specific context of AI-assisted writing). We list them in Table~\ref{table:2}.

\begin{table}[h!]
\centering
\renewcommand\theadfont{\bfseries}
\begin{tabular}{@{}>{\raggedright\arraybackslash}p{3cm}  % Procedural factors
                >{\raggedright\arraybackslash}p{3cm}     
                >{\raggedright\arraybackslash}p{3cm}    
                >{\raggedright\arraybackslash}p{3cm}@{}} 

\textbf{Factors} & \multicolumn{3}{c}{\textbf{Sources}} \\
\addlinespace
& \textit{\citet{li2024does}} & \emph{\citet{zhang2024secret}} & \emph{\citet{liao2023ai}}\\
\midrule
\textbf{Detectability} &  &  & \\
\midrule
\textbf{Regulatory penalty for misclaim} &  &  & \\
\midrule
\textbf{Perceived negative effects of disclosure} & Decrease in rating of writing quality & Perceived external judgment&  \\
\midrule
\textbf{Perceived positive effects of disclosure} &  &  & Transparency\\
\midrule
\textbf{Prevalence of disclosure} &  &  & \\
\bottomrule
\end{tabular}
\caption{Mapping social factors to sources}
\label{table:2}
\end{table}

\subsection{Personal Factors}
\label{sec:appendix-personal}

Only one previous work has studied the effect of a writer's personal factors on their disclosure decisions~\cite{zhang2024secret}. Specifically, \citet{zhang2024secret} considered demographics (including gender, age, AI use frequency, and education), self-efficacy of writing, and internal judgments on using AI for writing. In our list, we further include self-efficacy of using AI to assist writing and internal judgments of disclosure as potential factors that influence writers' disclosure behaviors.
% \mina{need to describe which demographic features they used (and what they found if appropriate)} However, it only considered \mina{instead of pointing it out as a flaw (also note that in the appendix without much context, it's hard to understand what we are describing), describe it as difference between the personal factors they considered and ours described in the main text} one dimension of both self-efficacy and internal judgments. Specifically, \citet{zhang2024secret} considered self-efficacy of writing and internal judgments on using AI for writing, while overlooking the potential influences of self-efficacy of using AI to assist writing and internal judgments of disclosure.

% \begin{table}[h!]
% \centering
% \renewcommand\theadfont{\bfseries}
% \begin{tabular}{@{}>{\raggedright\arraybackslash}p{3cm}  % Procedural factors
%                 >{\raggedright\arraybackslash}p{3cm}     
%                 >{\raggedright\arraybackslash}p{3cm}    
%                 >{\raggedright\arraybackslash}p{3cm}@{}} 

% \textbf{Personal factors} & \multicolumn{1}{c}{\textbf{Source}} \\
% \addlinespace
% & \emph{\citet{zhang2024secret}} \\
% \midrule
% \emph{Self-efficacy} &  self-efficacy (for writing only)\\
% \midrule
% \emph{Internal judgments} &  internal judgments (for using AI only)\\
% \midrule
% \emph{Demograhics} &  demograhics\\
% \bottomrule
% \end{tabular}
% \caption{Mapping personal factors to sources}
% \label{table:3}
% \end{table}

\end{document}